\documentclass[]{spie}  %>>> use for US letter paper
\usepackage{amsmath,amsfonts,amssymb}
\usepackage{graphicx}
\usepackage[colorlinks=true, allcolors=blue]{hyperref}
\usepackage{amssymb}
\usepackage{geometry}
\usepackage[misc]{ifsym}
\usepackage{bbding}
\usepackage{amsmath}
\usepackage{graphicx}
\usepackage{multirow}
\usepackage{array}

\usepackage{float}
\usepackage{multicol}

\title{SegmentAnything helps microscopy images based automatic and quantitative organoid detection and analysis}
\author[a, *]{Xiaodan Xing}
\author[b, *]{Chunling Tang}
\author[b]{Yunzhe Guo}
\author[c, d]{Nicholas Kurniawan}
\author[a, e]{Guang Yang}
\affil[a]{National Heart and Lung Institute, Imperial College London, London, UK}
\affil[b]{Centre for Craniofacial \& Regenerative Biology, King’s College London, UK}
\affil[c]{Department of Biomedical Engineering, Eindhoven University of Technology, Eindhoven, 
Netherlands}
\affil[d]{Institute for Complex Molecular Systems (ICMS), Eindhoven, Netherlands}
\affil[e]{Department of Bioengineering, Imperial College London, London, UK}

\authorinfo{*Send correspondence to X. Xing (xxing@imperial.ac.uk) and G. Yang (g.yang@imperial.ac.uk). Xiaodan and Chunling contributed equally to this paper.
}

\begin{document} 
\maketitle
%This pattern could be obtained by Doppler guide wire studies, which introduces inevitable radiation and possible complications. 
\begin{abstract}
Organoids are self-organized 3D cell clusters that closely mimic the architecture and function of in vivo tissues and organs. Quantification of organoid morphology helps in studying organ development, drug discovery, and toxicity assessment. Recent microscopy techniques provide a potent tool to acquire organoid morphology features, but manual image analysis remains a labor and time-intensive process. Thus, this paper proposes a comprehensive pipeline for microscopy analysis that leverages the SegmentAnything to precisely demarcate individual organoids. Additionally, we introduce a set of morphological properties, including perimeter, area, radius, non-smoothness, and non-circularity, allowing researchers to analyze the organoid structures quantitatively and automatically. To validate the effectiveness of our approach, we conducted tests on bright-field images of human induced pluripotent stem cells (iPSCs) derived neural-epithelial (NE) organoids. The results obtained from our automatic pipeline closely align with manual organoid detection and measurement, showcasing the capability of our proposed method in accelerating organoids morphology analysis.

\end{abstract}


\section{Description of purpose}
Organoids are self-organized 3D tissues typically derived from stem cells, exhibiting key functional, structural, and biological complexity similar to organs \cite{zhao2022organoids}. Their close biological resemblance makes organoid culture analysis crucial for advancing biological studies, as it aids in understanding the extent to which organoids resemble their in vivo counterparts.

The analysis of organoid morphology is commonly performed by capturing images of the organoids grown in multi-well plates. However, existing methods have limitations since they aggregate cell growth information over an entire well, rather than providing information about individual organoids and their constituent cells \cite{mukashyaka2023cellos}. Unfortunately, manually demarcating organoids in microscopy images poses significant challenges. The sheer number of organoids in a single whole slice microscopy image can reach thousands, making manual demarcation a laborious and time-consuming task.

To tackle this challenge, researchers have introduced deep learning algorithms, such as Startdist \cite{stardist} and Cellos \cite{mukashyaka2023cellos}. However, these deep learning-based approaches demand a substantial amount of annotated data for effective algorithm training. Moreover, their limited scope in handling various modalities hinders their generalizability. For each distinct type of microscopy image, re-training the models becomes necessary, posing practical limitations on their applicability.

\begin{figure}
    \centering
    \includegraphics[width=14 cm]{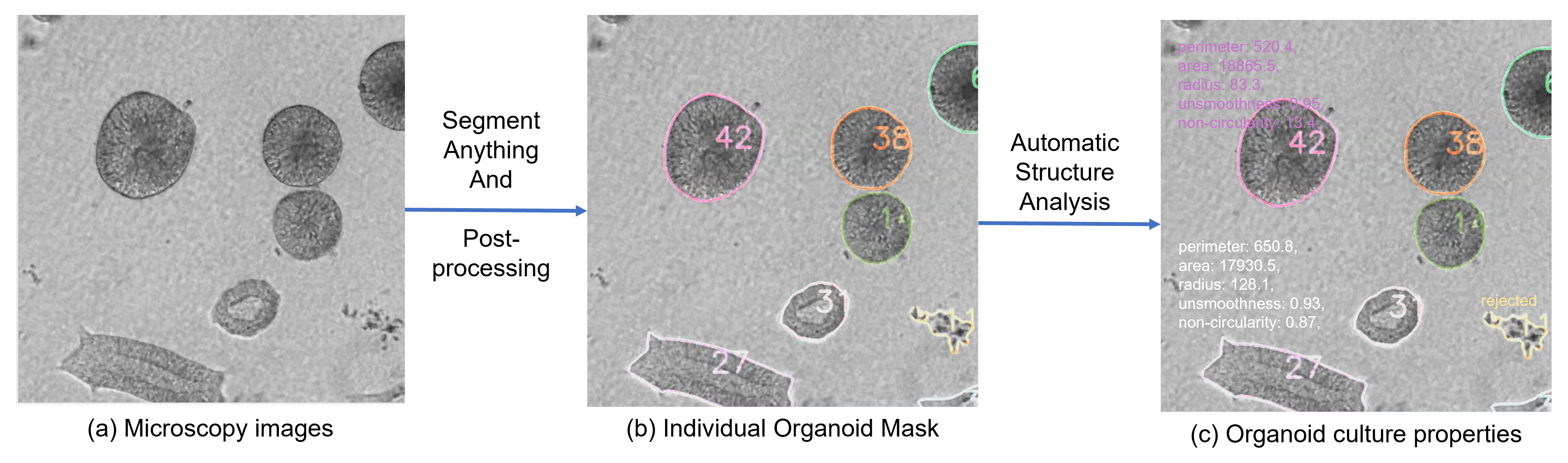}
    \caption{Overview of the proposed method.}
    \label{fig:flowchart}
\end{figure}

In this study, we explore the potential of SegmentAnything \cite{kirillov2023segment}, a foundation model trained on an extensive dataset of 11 million images encompassing diverse modalities, to automate individual organoid detection in microscopy images. Moreover, we have integrated comprehensive post-processing and analysis of morphological properties using the masks generated by SegmentAnything. The workflow is demonstrated in Fig. \ref{fig:flowchart}. Our main claim is that this proposed pipeline enables both automatic and accurate organoid detection, as well as fully automated organoid morphology analysis.

To further validate our hypothesis, we conducted a comparison between the outcomes of our research and those from a previously published peer-reviewed paper about hiSPCs derived NE organoids \cite{chunling}. Theses NE organoids are generated to mimic neural tube development at early embryogenesis stage. They are defined with round morphology but different size during culture in vitro. Briefly, the result from our study shows the robustness of SegmentAnything in detecting NE organoids from bright-field images. Furthermore, automatically organoids size quantification closely aligns with the manually measured results from the publication, reinforcing the efficacy of our proposed approach. We are the first one investigating the efficacy of SegmentAnything on organoid detection, and all codes are open sourced in \url{https://github.com/XiaodanXing/SAM4organoid}.

\section{Method}
\noindent \textbf{Data acquisition.} Bright-field images used in this paper were obtained under the protocol described in \cite{chunling}. The images were captured using Leica DMi8 microscope (Leica) equipped with 10×/0.32 objective lens. We obtained one whole slide image from each group. 

\begin{figure}
    \centering
    \includegraphics[width=9 cm]{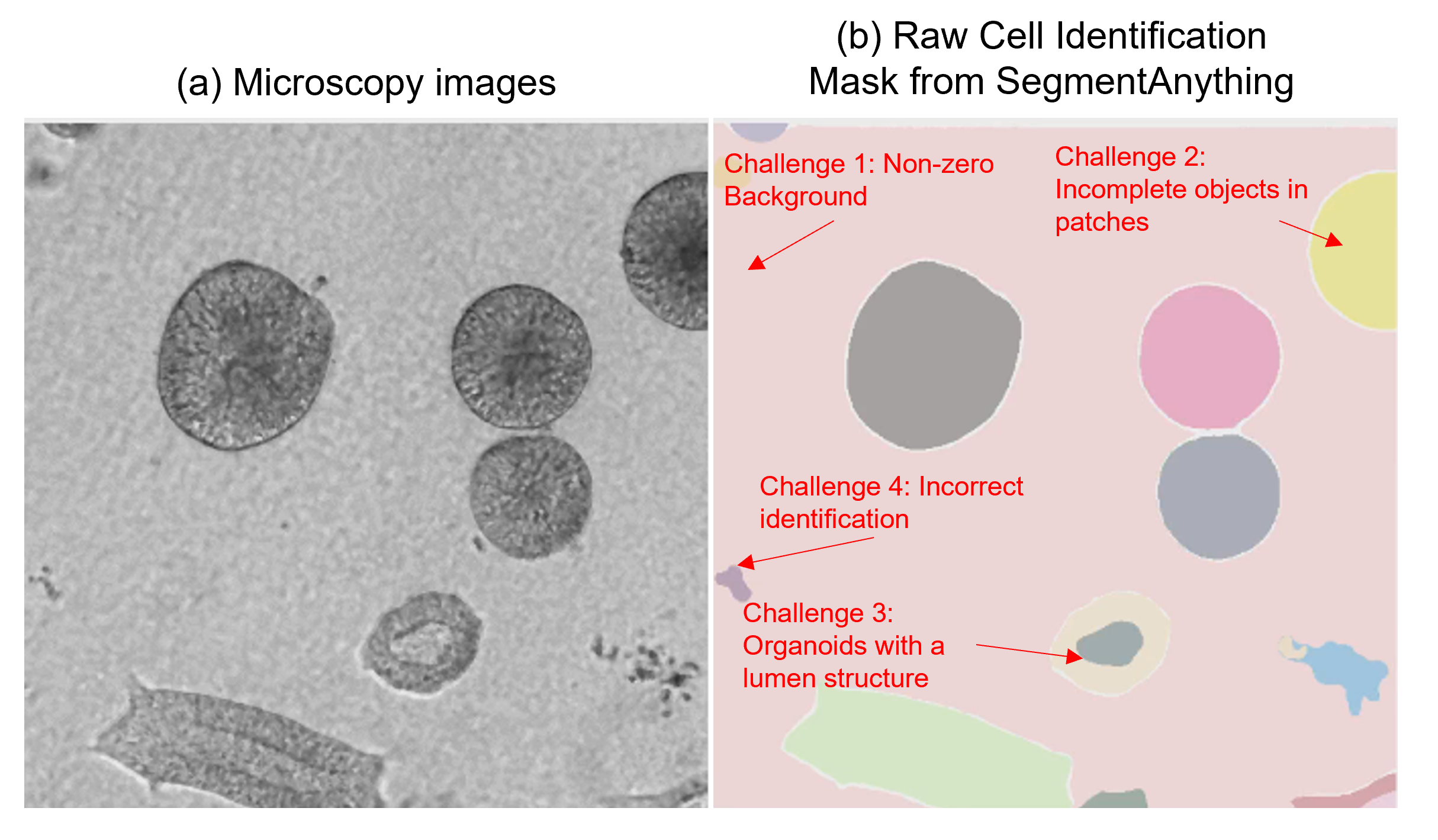}
    \caption{Challenges in directly applying SegmentAnything in real-world organoid morphology analysis workflow.}
    \label{fig:challenge}
\end{figure}

\noindent \textbf{SegmentAnything and post processing.} In our research, we utilized the Python API for SegmentAnything and evaluated three pretrained models \cite{kirillov2023segment}, namely ViT-B, ViT-H, and ViT-L, ultimately selecting the ViT-H model for inference due to its consistent performance across various microscopy analyses. However, we encountered challenges with the SegmentAnything-generated masks, as is shown in FIg. \ref{fig:challenge}, which required post-processing to achieve accurate cell identification.

The first issue we encountered was that SegmentAnything sometimes misidentified the background as an object, resulting in non-zero indices for the background in the masks. Secondly, the high resolution of whole microscopy images necessitated the use of cropped patches for model fitting. However, this approach introduced incomplete organoids along the edges of the patches, leading to erroneous analysis of morphological properties. To address these concerns, we implemented an automated process where we the boundaries of the image patches were examined, and all objects located in these regions were excluded. A third challenge was observed with organoids possessing a lumen structure, where the model inaccurately demarcated the regions into two separate objects. To rectify this problem, we computed the maximum boundary of each mask and unified all values within this boundary. Lastly, debris might be erroneously identified as objects (organoids in this scenario) by the model. Unfortunately, we have not yet found an automated method to remove them. Thus, we manually marked these non-organoid structures and deleted them, which, when compared to manually identifying all organoid structures, proved to be a relatively simpler task. 

\noindent\textbf{Property Analysis}: We conducted a comprehensive analysis of each organoid, computing five distinct properties to characterize their characteristics:
\vspace{-2mm}
\begin{enumerate}
\vspace{-1mm}
    \item Perimeter: This property quantifies the total length of the organoid's boundary, providing a measure of its overall shape complexity.
    \vspace{-1mm}
    \item Radius: To estimate the organoid's size, we calculated the average distance from the center of the cell to various points on its perimeter.
    \vspace{-1mm}
    \item Area: This property corresponds to the number of pixels encompassed within the organoid, serving as a direct indicator of its size.
    \vspace{-1mm}
    \item Non-smoothness: Non-smoothness reflects the local variation in radius lengths along the organoid boundary. A higher non-smoothness value indicates a more irregular and less smooth boundary. To compute this property, we fitted an ellipse to the organoid’s boundary and determined the smoothness as the ratio of perimeters between the fitted ellipse and the original contour.
    \vspace{-1mm}
    \item 	Non-circularity: We employed the following equation to evaluate the extent to which the organoid resembles a perfect circle:$$Non-circularity =|(Perimeter^2)/(4\pi \times Area)-1|$$
\end{enumerate}

\section{Results}

\begin{figure}
    \centering
    \includegraphics[width=15 cm]{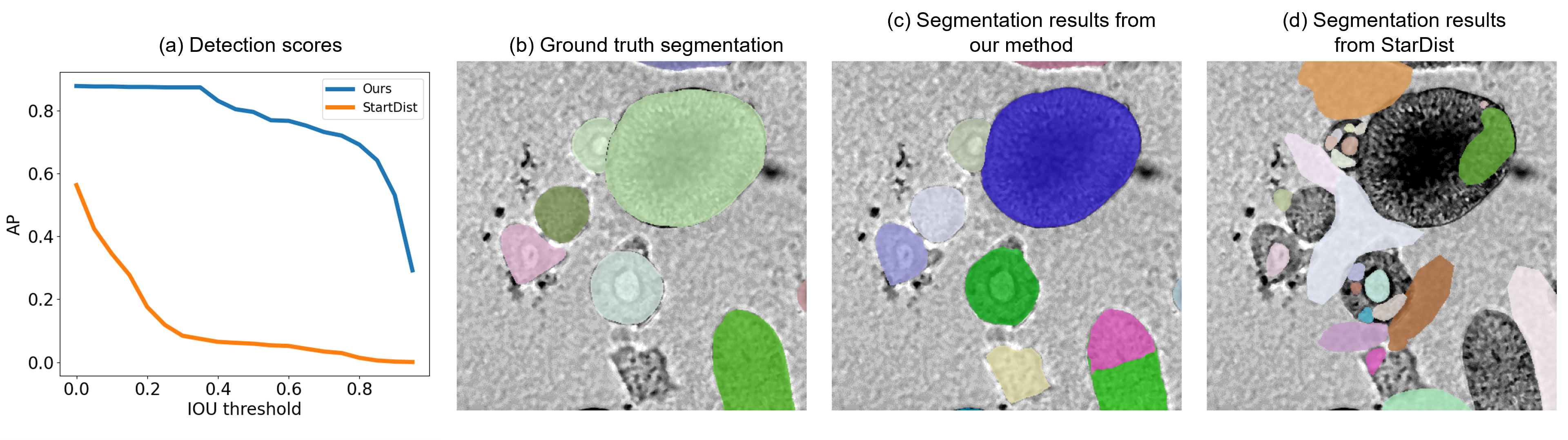}
    \caption{(a) The average detection scores comparison between our method and the StarDist. (c) and (d) are the segmentation results from our method and the StarDist algorithm, respectively. }
    \label{fig:iou}
\end{figure}

We conducted an analysis of bright-field (BF) images of NE organoids formed in the neural induction medium with 2\% and 8\% Geltrex respectively at day 7 and day 18. The findings are presented below. 

We conducted a comparison of the mean average precision (mAP) between the organoid detection results obtained from our method and those obtained from the open sourced StarDist method \cite{stardist} in Fig. \ref{fig:iou}. Instead of training the StarDist method from scratch, we inferenced the '2D\_versatile\_fluo' model with default settings. The mAP comparison results are depicted in Fig. 4(a), while the segmentation comparison results are presented in Fig. 4(c) and (d). To ensure a fair and unbiased comparison, we refrained from manually removing any wrongly segmented regions (as described in challenge 4) from our proposed method. The results clearly demonstrate that the StarDist method, without any training or fine-tuning on the test modality, failed to achieve accurate segmentation on organoids.
\begin{figure}
    \centering
    \includegraphics[width=15 cm]{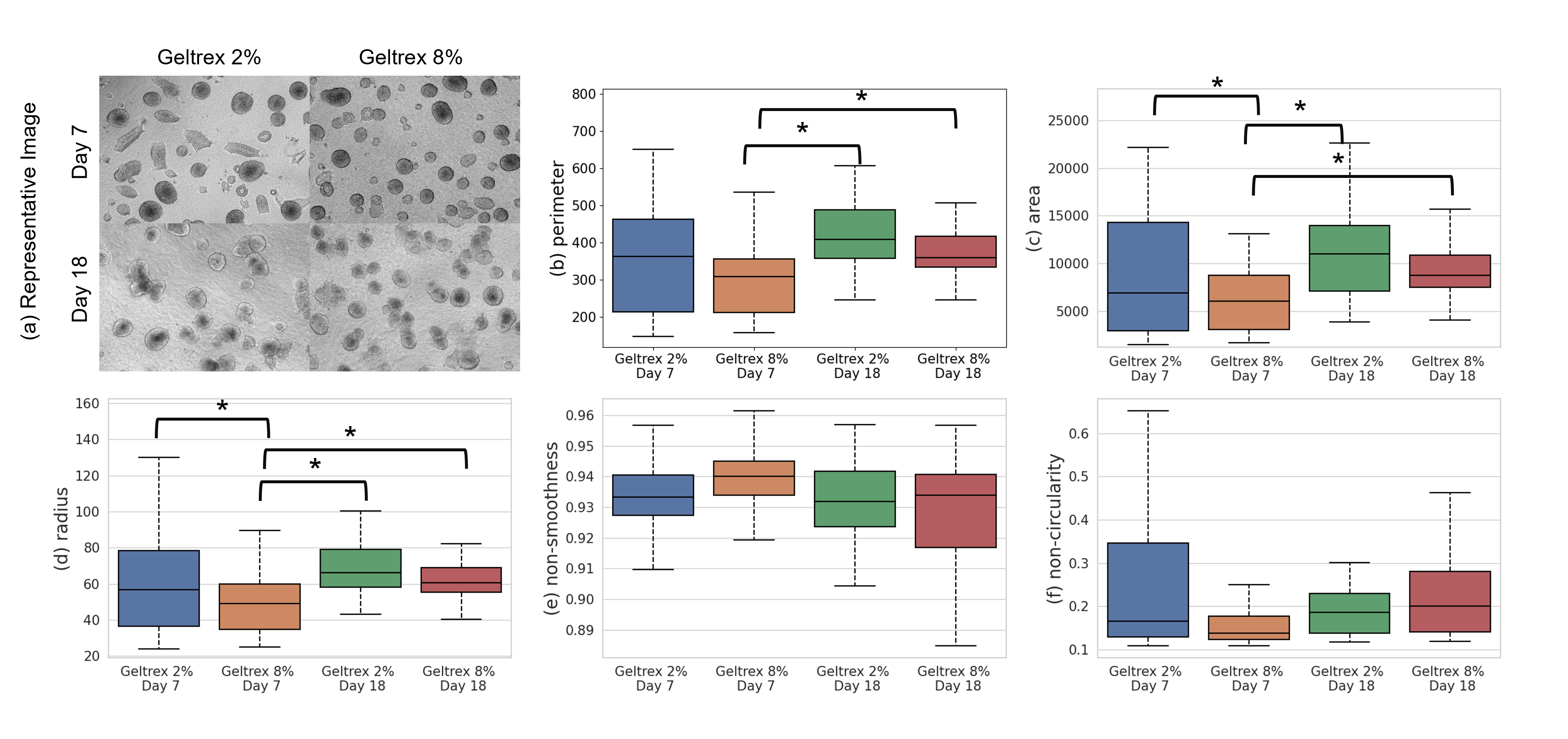}
    \caption{The analysis results among four groups. In (a) we presented the representative image patches from four different groups. * in figure (b-f) represents a significant difference using Student t-test. }
    \label{fig:size}
\end{figure}
We also analyzed the morphological features of organoids among different groups in Fig, \ref{fig:size}. Our results indicate that in the later stage of organoid formation (day 18), a higher concerntration of Geltrex leads to smaller organoid sizes, which aligns with the hypothesis that Geltrex, being a hydrogel, undergoes solidification at 37 degrees Celsius, thereby exerting pressure on organoid formation from the paper \cite{chunling}. Furthermore, our results are in agreement with the manually annotated results, highlighting the capability of our proposed toolbox in facilitating biological studies. The consistency between our automated analysis and the manually derived findings demonstrates the reliability and effectiveness of our approach in cellular analysis, offering valuable insights for further research and experimentation.

\section{Conclusions}
In this paper, we utilized the SegmentAnything model in automatic organoid structure identification in microscopy images. We claim that the SegmentAnything model showed promising performance, and our post-processing efforts were also necessary to enhance the accuracy of organoid structure detection and ensure reliable organoid morphology analysis. Overall, this research contributes to the field of organoid analysis in microscopy images by presenting an efficient approach for individual organoid detection and morphology analysis without any pre-requisites on data annotation. The automated pipeline offers promising avenues for accelerating and enhancing organoid features characterization and quantification, paving the way for further advancements in organoid research and related disciplines.\\

% % References
% \bibliography{report} % bibliography data in report.bib
% \bibliographystyle{spiebib} % makes bibtex use spiebib.bst
\bibliographystyle{spiebib}
\bibliography{bibtex.bib}
\end{document}